\documentclass[aps,prb,showpacs,groupedaddress,superscriptaddress,twocolumn]{revtex4}
\usepackage{graphicx}
\usepackage{epstopdf}
\usepackage{amsmath}
\usepackage{bm}
\usepackage{xcolor}
\usepackage{subfigure}
\begin{document}
\title{Crossed Andreev effects in two-dimensional quantum Hall systems}

\author{Zhe Hou}
\affiliation{International Center for Quantum Materials, School of Physics, Peking University, Beijing 100871, China}
\author{Yanxia Xing}
\affiliation{Department of Physics, Beijing Institute of Technology, Beijing 100081, China}
\author{Ai-Min Guo}
\affiliation{Department of Physics, Harbin Institute of Technology, Harbin 150001, China}
\author{Qing-Feng Sun}
\email[]{sunqf@pku.edu.cn}
\affiliation{International Center for Quantum Materials, School of Physics, Peking University, Beijing 100871, China}
\affiliation{Collaborative Innovation Center of Quantum Matter, Beijing 100871, China}

\begin{abstract}
We study the crossed Andreev effects in two-dimensional conductor/superconductor hybrid systems under a perpendicular magnetic field.
Both graphene/superconductor hybrid system and electron gas/superconductor one are considered.
It is shown that an exclusive crossed Andreev reflection, with other Andreev reflections being completely suppressed,
is obtained in a high magnetic field because of the chiral edge states in the quantum Hall regime.
Importantly, the exclusive crossed Andreev reflection not only holds for a wide range of system parameters,
e.g., the size of system, the width of central superconductor,
and the quality of coupling between the graphene and the superconductor,
but also is very robust against disorder. When the applied bias is within the superconductor gap, a perfect Cooper-pair splitting process with high-efficiency can be realized in this system.
\end{abstract}


\maketitle

\section{introduction}
The Cooper pairs in a superconductor can be used as a natural source of nonlocal Einstein-Podolsky-Rosen (EPR) electron pairs.\cite{EPR1,EPR2} By splitting the electrons in a Cooper pair, one obtains two spatially separated electrons which still keep their spin and momentum entangled. The Cooper pair splitting is the inverse process of crossed Andreev reflection (CAR) which has been extensively studied for many years.\cite{CAR1,CAR2,CAR3,CAR4,CPS1,CPS2,quantum information,add0,add1,adddot,add3,add4,add5,CPS3,add51,addgraphene, referee, referee1} The CAR process occurs at the interface between a normal conductor and the superconductor, where an electron is injected from one terminal of the normal conductor and is then reflected out as a hole at the other terminal, and a Cooper pair forms in the superconductor. Except for the CAR, there also exists local Andreev reflection (LAR), where the hole is reflected into the same terminal. The CAR process usually competes with the LAR one. To split a Cooper pair efficiently, the LAR process has to be suppressed.

Up to now, many proposals have been put forward to realize the splitting of Cooper
pairs\cite{CAR1,CAR2,CAR3,CAR4,CPS1,CPS2,quantum information,add0,add1,adddot,add3,add4,add5,CPS3,add51,addgraphene}
for its promising applications in quantum communication
and quantum computing.\cite{quantum information,quantum information2,quantum computing}
For example, some Cooper-pair splitters have been theoretically proposed in the system by coupling a superconductor with quantum dots,\cite{quantum information,add1,adddot}
carbon nanotubes,\cite{add3,add4} Luttinger liquid wires,\cite{add5} and graphene.\cite{CPS3,add51,addgraphene}
On the experimental sides,\cite{CPS4,add2,CPS5} the Cooper-pair splitters have been realized in the system by coupling
a superconductor with two quantum dots, where a central superconducting finger is connected
with two quantum dots and each quantum dot is coupled with a metallic lead.
In the Coulomb blockade regime, the electrons in a Cooper pair can tunnel into different
leads coherently from the superconductor, and the LAR process can be considerably suppressed by tuning the energy levels of the
quantum dots. However, in this Cooper-pair splitter, the LAR process cannot
be completely suppressed, and thus the entangled electrons are not spatially separated
completely.\cite{CPS1} Furthermore, to improve the efficiency of the CAR process,
the parameters, such as the bias and the gate voltage of the quantum dots, have to be accurate. And it may be difficult for experimental implementation of the Cooper-pair splitter.

Recently, due to the emergence of topological insulators and Majorana Fermion,
some Cooper-pair splitters are proposed based on the hybrid system of the superconductor and
the topological insulators or the Majorana Fermion.\cite{CPS1,addTI,addMF,add60,add61,add6}
For example, in a two-dimensional (2D) topological insulator-superconductor-2D topological insulator junction,
an all-electric Cooper-pair splitter was proposed by Reinthaler
\emph{et al.}.\cite{add6}
James \emph{et al.}\cite{CPS1} proposed an exclusive CAR device by inducing superconductivity
on a AIII class topological insulator wire which supports two topological phases with one or
two Majorana fermion end states. In the phase with two Majorana fermions, the LAR is completely
suppressed at the normal lead/topological superconductor interface at zero bias,
resulting in correlated and spin-polarized currents in the leads.

However, all of the previous Cooper-pair splitters have several disadvantages.
Since the incoming electron and the outgoing hole in the CAR process locate in spatially separated terminals,
the width $L$ of the central superconductor is required to be less than the superconducting
coherent length $\xi$, and the CAR coefficient would decay quickly to zero when $L>\xi$.
In usual, the LAR also occurs inevitably, in which the outgoing hole comes back to
the same terminal as the incoming electron. In many Cooper-pair splitters,
the LAR is usually much larger than the CAR and the efficiency of the Cooper-pair splitting
is quite low.
Furthermore, many Cooper-pair splitters are too elaborate to be realized experimentally. And they can work only under certain special parameters, and are usually not robust against
disorders and impurities, which exist inevitably in the experiments. As a result, the CAR process is
strongly suppressed and the Cooper-pair splitting efficiency is very low.
Here, we also notice that very recently, Zhang \emph{et al.} proposed a Cooper-pair splitter
based on a quantum anomalous Hall insulator (QAHI).\cite{add7} Due to
the unidirectionality of the chiral edge states in the QAHI, the LAR can be suppressed completely
and only the CAR occurs in the QAHI-superconductor-QAHI junction. Consequently, this QAHI-based Cooper-pair splitter can be very efficient and be robust against the
disorders, and can work even if the size of the superconductor electrode is much larger
than the superconducting coherent length. However, it is very difficult to fabricate the QAHI in the experiment,
although the QAHI has been successfully realized in the magnetic topological insulator
with the temperature being at the order of mK.\cite{add8,add81}
Therefore, it is an urgent task to propose a perfect Cooper-pair splitter which is of high-efficiency and is
robust against the disorder, and works in large-superconductor size.

Graphene is a 2D material with a unique band structure.\cite{graphene,graphene2}
Electrons in graphene exhibit relativistic-like behavior near the Dirac point.
One of the peculiar properties of graphene is the half integer quantum Hall effect
with the chiral edge states.
In this paper, we investigate the electron transport through a three-terminal
graphene/superconductor system, and propose a perfect Cooper-pair splitter
with the aid of the chiral edge states in the quantum Hall system.
In general, the quantum Hall effect and the superconductivity are mutually exclusive,
because the former phenomenon exists in the presence of strong magnetic field, whereas the latter one will be
destroyed by the strong magnetic field.
However, with modern progress in materials science,
the quantum Hall effect can be observed at much smaller magnetic field, which ensures
the possibility of coexistence of the quantum Hall effect and the superconductivity.\cite{interplay1,interplay2,interplay3,interplay4,interplay5}
For example, both the quantum Hall effect and the superconductivity have been successfully realized in the junction which
consists of the 2D electron gas and the Nb compounds, where
the 2D electron gas with high mobility possesses the quantum Hall regime under sufficiently small magnetic field and the Nb compounds have a high critical magnetic field.\cite{interplay3}
The coexistence of the quantum Hall effect and the superconductivity
has also been observed in the graphene/superconductor hybrid system.\cite{interplay4}

By using the tight-binding model
and the non-equilibrium Green's function method, we obtain expressions of Andreev
reflection coefficients and normal transmission coefficients under different magnetic fields.
In strong magnetic field, the chiral edge states form in the graphene, and
the electrons (holes) can be reflected unidirectionally as
the holes (electrons) at the interface between the graphene and the superconductor.
Because of the unidirectionality of the chiral edge states, the outgoing hole will be transmitted to the other graphene
terminal and only the CAR occurs, as shown in Fig.1(a).
Notice that the LAR happens only if the outgoing hole is scattered from one edge of the graphene to the other edge.
In the quantum Hall regime, the scattering between the two edges is almost impossible, and hence the LAR is completely inhibited and an exclusive CAR emerges.
Fig.2 shows the main results in this paper, where only the CAR coefficient $T_{13A}$ has large value
and the other Andreev reflection coefficients are almost zero.
As a result, this device can serve as a perfect Cooper-pair splitter with high-efficiency.
This Cooper-pair splitter can be very robust against the disorder. As long as the chiral edge states are present, the Cooper-pair splitter can work well.
Furthermore, it works well even when the width $L$ of the superconductor is larger than the superconducting
coherent length $\xi$ and the width $L$ breaks through the size limitation of previous
Cooper-pair splitters.
In addition, the exclusive CAR process, with the LAR being completely suppressed, can hold for
a wide range of system parameters, such as the width of the graphene nanoribbon,
the coupling between the superconductor and the graphene.
Finally, a 2D electron gas is considered instead of the graphene and similar results are
obtained due to the emergence of the chiral edge states, which are induced by external magnetic field (see Fig.1(b)).

The rest of the paper is constructed as follows. In Sec.II,
the theoretical model is presented and the expressions of the normal transmission coefficients
and the Andreev reflection coefficients are derived. In Sec.III, we numerically investigate
the transmission coefficient, the CAR coefficient, and the LAR coefficient, and discuss the characteristics of the proposed
Cooper-pair splitter in the graphene/superconductor hybrid system. In Sec.IV, we change the Dirac point of the
graphene to demonstrate the regime of the CAR and the LAR. Finally, the results are summarized in Sec.V. The Cooper-pair splitter
in the 2D electron gas/superconductor hybrid system is presented in Appendix.

\begin{figure}
\includegraphics[width=4.1cm,totalheight=2.3cm, clip=]{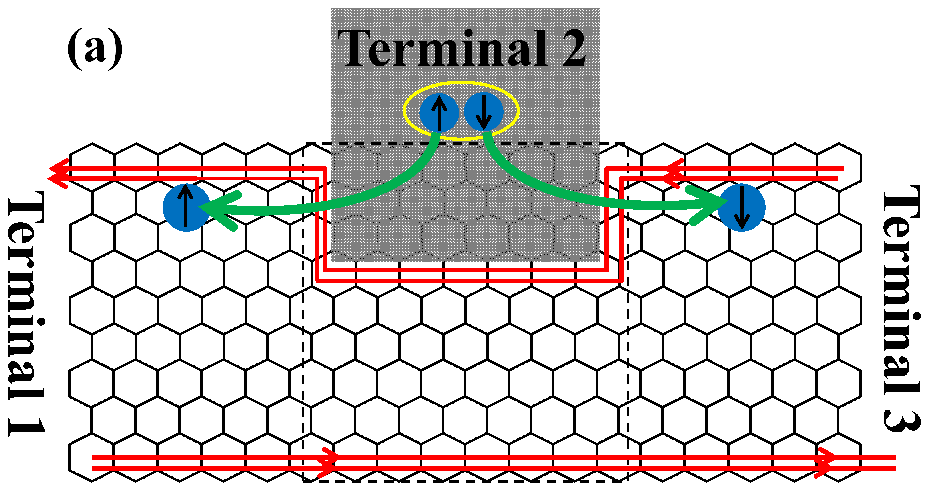}
\includegraphics[width=4.1cm,totalheight=2.3cm, clip=]{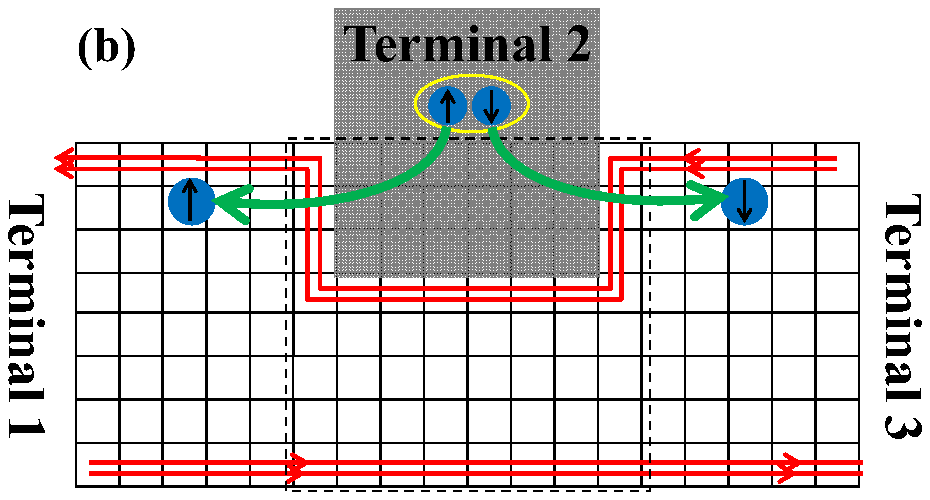}
\caption{ (Color online) Schematic diagram for three-terminal hybrid system of (a) graphene/superconductor
and (b) two-dimensional electron gas/superconductor.
The red lines denote the edge states due to the strong magnetic field $B$.
With the aid of the unidirectional chiral edge states, the Cooper pairs in the superconductor lead can be split into two separated terminals.}
\end{figure}

\begin{figure}
\centering
\includegraphics[width=7.0cm,totalheight=5.3cm, clip=]{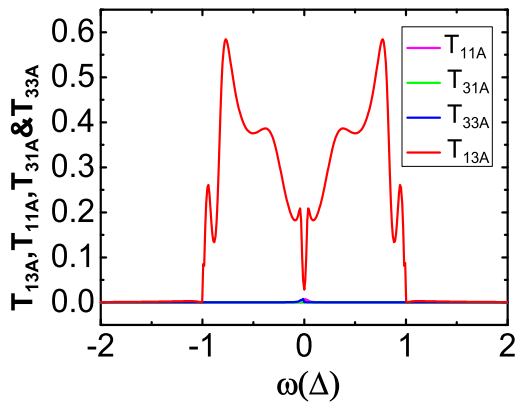}
\caption{
(Color online) Andreev reflection coefficients $T_A$ versus
incident energy $\omega$ for the graphene/superconductor hybrid system at $\phi=0.003$.
Due to the chiral edge states, the exclusive CAR coefficient $T_{13A}$ is very large, with the other Andreev reflection coefficients being
prohibited.
The parameters are listed as follows:
the Dirac point energy $E_0=-5\Delta$, the width of the graphene nanoribbon $N=50$,
the length of the central region $L=30$, the size of the covered area of the superconductor $M=15$,
the coupling strength between the graphene and the superconductor $t_c=t$, and the disorder strength $W$=0.}
\end{figure}

\section{model and formulism}

We consider a three-terminal system by coupling a zigzag edged graphene nanoribbon
with a superconductor lead (as shown in Fig.1(a)). The central region with width $N$ and length $L$, as labeled by the rectangular area in Fig.1(a),
is partly covered by the superconductor lead. The covered region is described by
width $M$ and length $L$ with $2M(2L-1)$ carbon atoms.
In Fig.1(a), it shows a system with $N=5$, $L=7$, and $M=2$.

The Hamiltonian of the system is
\begin{eqnarray}
H=H_G+H_S+H_C,
\end{eqnarray}
where $H_{\rm G}$, $H_{\rm S}$, and $H_{\rm C}$ are the Hamiltonians of the graphene nanoribbon,
the superconductor lead, and the coupling between them, respectively.
In the tight-binding representation, $H_{\rm G}$ is:\cite{HG1,HG2}
\begin{eqnarray}
H_{\rm G}=\sum_{i,\sigma}E_ia_{i\sigma}^{\dagger}a_{i\sigma}-
\sum_{\langle ij\rangle,\sigma}t{\rm e}^{{\rm i}\phi_{ij}}a_{i\sigma}^{\dagger}a_{j\sigma},
\end{eqnarray}
where $a_{i\sigma}^{\dagger}$ and $a_{i\sigma}$ are the creation and annihilation operators
at the discrete site $i$, and $E_i=E_0+\omega_i$ denotes the on-site energy.
$E_0$ is the Dirac point energy which can be controlled experimentally by gate voltage
and $\omega_i$ is the on-site disordered energy.
In the experiment, disorder and impurity exist inevitably.
The disorder in graphene p-n junction can result in several extra conductance plateaus.\cite{HG2,disorder1,disorder2} Furthermore,
the charge puddle disorder has been confirmed in the graphene by recent experiments.\cite{pud1,pud2}
Here, we consider that the disorder only exists in the central scattering region.
$\omega_i=0$ at the left and right graphene terminals; while for the central region,
$\omega_i$ is uniformly distributed in the range $[-W/2, W/2]$ with $W$ being the disorder strength.
The second term in $H_G$ describes the nearest-neighbor hopping.
Because of a uniform perpendicular magnetic field on the graphene nanoribbon, a phase
$\phi_{ij}$ is added in the hopping element and $\phi_{ij}=\int_i^j\vec{A}\cdot{\rm d}\vec{l}/{\phi_0}$ with the vector potential $\vec{A}=(-By,0,0)$
and $\phi_0=\hbar/e$. Note that in the superconductor lead and the covered graphene region,
no magnetic field exists due to the Meissner effect and $\phi_{ij}=0$ in these regions.

As for the superconductor lead, we consider the BCS Hamiltonian and $H_{\rm S}$ is:
\begin{eqnarray}
H_{\rm S}=\sum_{\bf{k},\sigma}\epsilon_{\bf{k}}b_{\bf{k}\sigma}^{\dagger}b_{\bf{k}\sigma}
+\sum_{\bf{k}}\left({\Delta}b_{\bf{k}\uparrow}^{\dagger}b_{\bf{-k}\downarrow}^{\dagger}+
{\Delta}b_{\bf{-k}\downarrow}b_{\bf{k}\uparrow}\right),
\end{eqnarray}
where $\Delta$ is the superconductor gap and $b_{\bf{k}\sigma}^{\dagger}(b_{\bf{k}\sigma})$ is the creation (annihilation) operator in the superconductor lead with momentum ${\bf k}=(k_x,k_y)$.
The coupling between the superconductor and the graphene is described by the Hamiltonian:
\begin {eqnarray}
H_{\rm C}=\sum_{i,\sigma}t_{\rm c}a_{i\sigma}^{\dagger}b_{i\sigma}+\rm{h}_{\cdot}\rm{c}_{\cdot},
\end{eqnarray}
where $b_{i\sigma}$ is the annihilation operator at site $i$ and
$ b_{i\sigma}=\sum_{\bf{k}}{\rm e}^{{\rm i}{\bf k}\cdot{\bf r}_i}b_{\bf{k}\sigma}$.
Here, ${\bf r}_i$ is the position of the $i$th carbon atom in real space and $t_c$ is
the coupling parameter between the superconductor and the graphene.

By using the non-equilibrium Green's function method, we can obtain the normal transmission coefficient $T_{nm(m\neq n)N}$, the CAR coefficient $T_{nm(m\neq n)A}$, and the LAR coefficient $T_{nm(m=n)A}$ between the graphene terminals $m=1,3$ and $n=1,3$:\cite{Ssurface,addTA1,addTA2}
\begin{eqnarray}
 T^\sigma_{nm(m\neq n)N}(\omega)&=&{\rm Tr}\left[{\bf G}^r{\bf \Gamma}_{n,\sigma}{\bf G}^a{\bf \Gamma}_{m,\sigma}\right],\\
 T^\sigma_{nmA}(\omega) &=&{\rm Tr}\left[{\bf G}^r{\bf \Gamma}_{n,{\bar\sigma}} {\bf G}^a{\bf \Gamma}_{m,\sigma}\right],
\end{eqnarray}
where $\sigma$ represents spin-up electron ($\uparrow$) and spin-down hole ($\downarrow$) in the $2\times2$ Nambu space, and ${\bar\sigma}=\uparrow(\downarrow)$ for $\sigma=\downarrow(\uparrow)$. Since the Pauli matrices $\sigma_{x,y,z}$ are commutative with the Hamiltonian $H$,
the transmission coefficient and the Andreev reflection one satisfy
$T^{\uparrow}_{nmN} =T^{\downarrow}_{nmN}$ and $T^{\uparrow}_{nmA} =T^{\downarrow}_{nmA}$,
and we ignore the superscript ``$\sigma=\uparrow,\downarrow$" in the following for representation simplicity.
$T_{13N}$ ($T_{31N}$) represents normal tunneling possibility from terminal-3 (terminal-1) to
terminal-1 (terminal-3). The LAR coefficient $T_{mmA}$ (CAR coefficient $T_{nm(m\neq n)A}$) is the probability
of an electron coming from the graphene terminal-$m$ and getting Andreev reflected as a hole into
the same terminal-$m$ (different terminal-$n$).
The linewidth function ${\bf {\Gamma}}_{n}(\omega)\equiv {\rm i} [{\bf {\Sigma}}_{n}^r -({\bf {\Sigma}}_{n}^r)^{\dagger}]$
and ${\bf {\Sigma}}_{n}^r(\omega)$ is the retarded self-energy due to the coupling between the central region
and the terminal-$n$. ${\bf {G}}^{r(a)} (\omega)$ is the retarded (advanced) Green's function of the central region
in the Nambu space, and ${\bf {G}}^{r} (\omega)=[{\bf {G}}^{a} (\omega)]^{\dagger} = (\omega {\bf I}-{\bf H}_{\rm C}-\sum_{n=1,2,3} {\bf {\Sigma}}_{n}^r(\omega) )^{-1}$ with ${\bf H}_{\rm C}$
being the Hamiltonian matrix for the central region.
For the self-energies ${\bf {\Sigma}}_{1(3)}^r(\omega)$ of the left and right leads,
we have ${\bf {\Sigma}}_{1(3),ij}^r (\omega)=t{\bf g}_{1(3),ij} ^r (\omega) t$,
where ${\bf g}_{1(3),ij} ^r (\omega)$ is the surface Green's function of terminal-1 (terminal-3)
which can be numerically calculated.\cite{surface} For the self-energy of the superconductor terminal,
we have ${\bf {\Sigma}}_{2,ij} ^r (\omega)=t_c {\bf g}_{2,ij} ^r (\omega) t_c$ and ${\bf g}_{2,ij} ^r (\omega)$ is:\cite{Ssurface}
\begin{eqnarray}
{\bf g}_{2,ij} ^r (\omega)&=&-{\rm i}\pi \rho \beta (\omega) J_0 (k_F |{\bf r}_i -{\bf r}_j|)
 \nonumber \\
&& \bigotimes
\left(
\begin{array}{cc}
1 & \Delta/\omega \\
\Delta/\omega  & 1
\end{array}
\right),
\end{eqnarray}
where $\rho$ represents normal density of states for the superconductor and
$J_0 (k_F |{\bf r}_i -{\bf r}_j|)$ is the 0th order Bessel function with $k_F$ being the Fermi wave
vector. $\beta (\omega)=-{\rm i} \omega / \sqrt {{\Delta}^2 -{\omega}^2 }$
for $|\omega|<\Delta$ and $\beta (\omega)=|\omega| / \sqrt {{\omega}^2 -{\Delta}^2 }$
for $|\omega|>\Delta$.\cite{Ssurface,addTA1,addTA2}
For simplicity, we assume that $J_0 (k_F |{\bf r}_i -{\bf r}_j|) =1$ for $i=j$ and otherwise
$J_0 (k_F |{\bf r}_i -{\bf r}_j|)=0$ for $i\not=j$. This assumption is reasonable because $k_F$ is usually in the order of ${{\AA}}^{-1}$.
After this assumption, the superconductor lead seems to be made up of one dimensional wires
and each carbon atom in the central region connects independently with a superconductor,
that has been assumed in Ref.[\onlinecite{1DSuperconductor}].
Then, Eq.(7) can be reduced as:
\begin{eqnarray}
{\bf g}_{2,ij} ^r (\omega)&=&-{\rm i}\pi \rho \beta (\omega) {\delta}_{ij} \bigotimes
\left(
\begin{array}{cc}
1 & \Delta/\omega \\
\Delta/\omega  & 1
\end{array}
\right).
\end{eqnarray}
By employing these transmission coefficients, the
current flowing from terminal 3 into the central region can be obtained straightforwardly:\cite{Ssurface,Chengshuguang}
\begin{eqnarray}
I_{3}&=&\frac{2e}{h}\int{\rm d}\omega \{\ T_{23N} (f_{3+}-f_{2})+T_{13N}(f_{3+}-f_{1+})\nonumber \\
&&+T_{13A}(f_{3+}-f_{1-})+T_{33A}(f_{3+}-f_{3-}) \}\
\end{eqnarray}
Here, $f_{\alpha \pm}(\omega)=1/ \{\ {\rm {exp}}[(\omega\mp eV_{\alpha})/k_B T]+1 \}\ $
is the Fermi distribution function for the terminal $\alpha$, with the temperature $T$
and the bias $V_{\alpha}$. The bias of the superconductor lead $V_{2}$ is set to zero.

In the following numerical calculations, we set the hopping energy $t=2.75$eV,
the nearest-neighbor carbon-carbon distance $a=0.142$nm, and the superconductor
gap $\Delta$=1meV. The magnetic field $B$ is expressed in terms
of $\phi\equiv (3 \sqrt{3} / 4)a^2 B/{\phi_0}$ and $(3 \sqrt{3} / 2)a^2 B$ is the magnetic
flux in the honeycomb lattice. In the presence of disorder,
the curves are averaged over 500 random configurations.

\section{three-terminal graphene/superconductor hybrid system}

\begin{figure}
\centering
\includegraphics[width=8.7cm,totalheight=7.5cm, clip=]{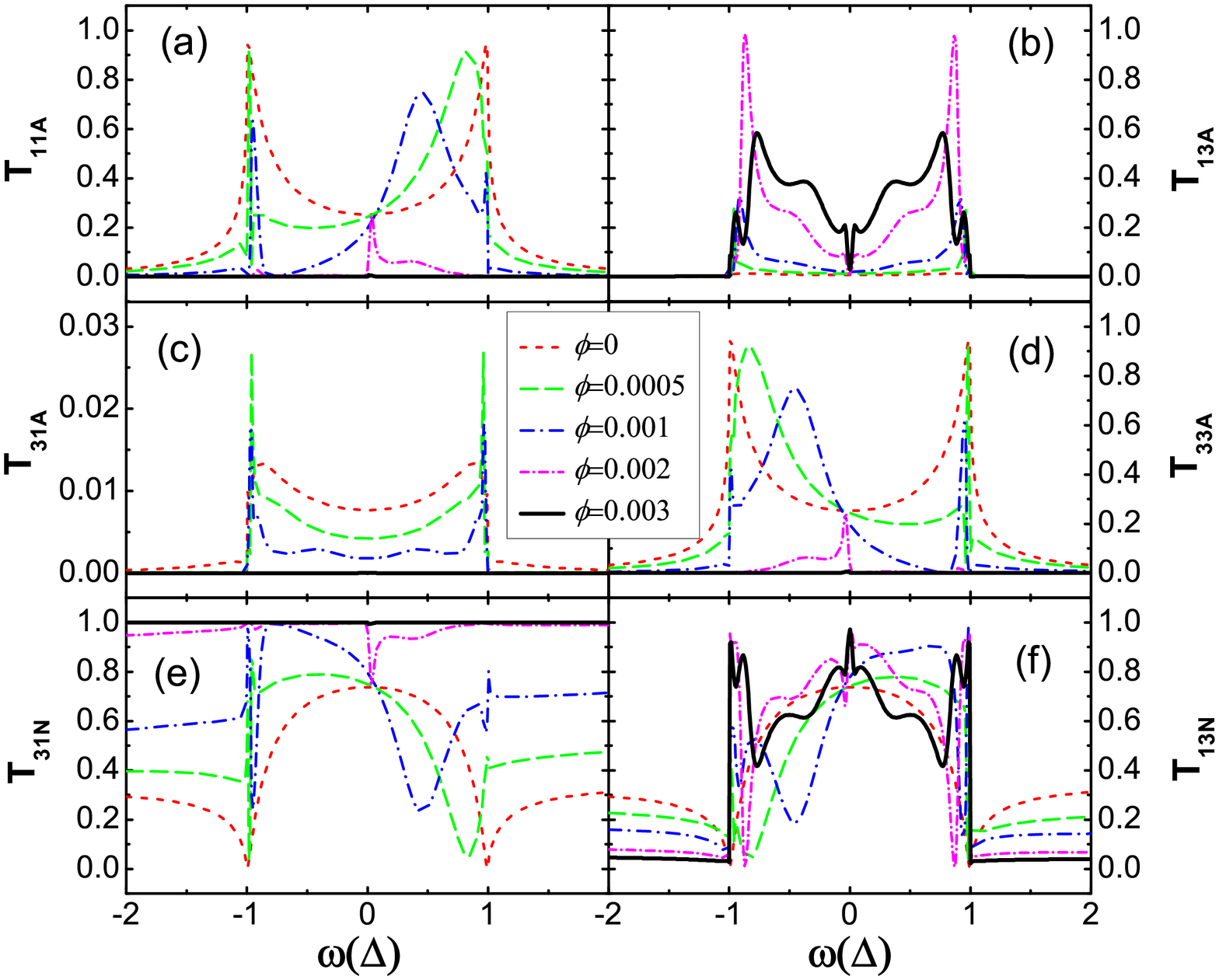}
\caption{
(Color online) Andreev reflection coefficients $T_A$ and normal transmission coefficients $T_{N}$ versus
incident energy $\omega$ for the graphene/superconductor hybrid system under different magnetic fields $\phi$ .
The other parameters are the same as those of Fig.2.
}
\end{figure}

We first study the electron transport properties of the graphene/superconductor
hybrid system under different magnetic fields. Fig.3 shows the Andreev reflection coefficients $T_A$ and
the normal transmission coefficients $T_{N}$ versus the energy $\omega$ of the incident electron.
In the absence of magnetic field $(\phi=0)$ or for weak magnetic field ($\phi=0.0005$ and $0.001$),
both the LAR process and the CAR one can occur for the incident electron from terminal-1 or from terminal-3 (see Figs.3(a)-3(d)).
The LAR coefficients $T_{11A}$ and $T_{33A}$ can be quite large, and the CAR coefficients $T_{31A}$
and $T_{13A}$ are relatively small when the energy $\omega$ is within the superconductor gap, i.e., $|\omega| < \Delta$.
Both the LAR and CAR coefficients decay quickly when $\omega$ is beyond the superconductor gap,
which is similar to usual normal-superconductor junction.\cite{normal-S}
The normal transmission coefficients $T_{13N}$ and $T_{31N}$ are large when $|\omega| < \Delta$ and
are decreased when $|\omega|>\Delta$,
because the tunneling from the graphene to the superconductor can happen when $|\omega|>\Delta$.
When the magnetic field is gradually increased to $\phi=0.003$,
one notices the following features.
(i) For the incident electron from terminal-1, the Andreev reflection coefficients $T_{11A}$ and $T_{31A}$ are
gradually declined to zero, and the normal transmission coefficient $T_{31N}$ is gradually increased
to one (see Figs.3(a), 3(c), and 3(e)). This indicates that the electrons tunnel directly from terminal-1 into terminal-3
without being Andreev reflected by the superconductor.
(ii) For the incident electron from terminal-3, the LAR coefficient $T_{33A}$ is shrunk to zero
and the CAR coefficient $T_{13A}$ is increased to a remarkable value (see Figs.3(b) and 3(d)),
and $T_{13A} +T_{13N} =1$ when $|\omega|<\Delta$ (see Figs.3(b) and 3(f)). This implies that when the electrons encounter the superconductor, they get Andreev reflected and no backscattering occurs.
It should be mentioned that the corresponding magnetic field $B$ is about 75 Tesla when $\phi=0.003$.
Since the width of the graphene nanoribbon in Fig.3 is very narrow (when $N=50$, the width is about $(3N-1)a \approx 21$nm), the magnetic field should be sufficiently strong so that the quantum Hall effect could appear.
While for a wide graphene nanoribbon, the magnetic field can be much weaker.

Now we explain the numerical results in Fig.3. In the weak or zero magnetic field,
the wave function of the incident electron can extend over the whole bulk of the graphene nanoribbon.
When the electron encounters the superconductor, the Andreev reflections occur at the interface of the graphene and the superconductor and give rise to either the LAR or the CAR.
Since the Fermi energy of the superconducting lead is $E_F=0$, which is far away from
the Dirac point $E_0=-5\Delta$ of the graphene in Fig.3,
the incident electron and the reflected hole locate in the same band (see Fig.8(c)). As a result,
the Andreev retro-reflection dominates in this situation and
the LAR is more pronounced than the CAR.\cite{Chengshuguang}
This accounts for the phenomenon that the LAR coefficient is much larger than the CAR one in the weak magnetic field.
While in the relatively strong magnetic field (e.g., $\phi=0.003$), both the Landau levels and the
edge states form, as depicted by the red lines in Fig.1.
In this case, the incident electron from terminal-1 transports along the bottom red line of Fig.1(a) and tunnels
directly into terminal-3 without encountering the superconductor. Thus, no backscattering and Andreev reflections occur, and $T_{11A}=T_{31A}=0$ and $T_{31N} =1$ which equals to the filling factor $\nu$ of the Landau levels. On the other hand,
the incident electron from terminal-3 transports along the top red line of Fig.1(a) and will be Andreev reflected at
the interface of the graphene and the superconductor.
Since the reflected hole lies in the same band as the incident electron mentioned above and their edge states have the same chirality, the reflected hole moves along the same direction of the incident electron. Then, only the CAR process occurs and the LAR process is completely suppressed.
Additionally, since the normal tunneling from the graphene into the superconductor is prohibited when the energy $\omega$ is within the superconductor gap, and the backscattering at the interface is also prohibited due to the chiral edge states, we have $T_{13A}+T_{13N}=1$ because of current conservation.

\begin{figure}
\includegraphics[width=8.7cm,totalheight=4cm, clip=]{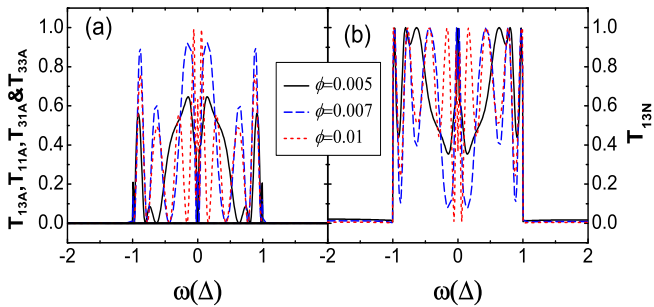}
\caption{
(Color online) (a) Andreev reflection coefficients $T_A$ and (b) transmission coefficient $T_{13N}$
versus incident energy $\omega$ in the regime of strong magnetic fields $\phi$.
In panel (a), only $T_{13A}$ has large value; $T_{11A}$, $T_{31A}$ and $T_{33A}$ are almost zero, and their curves overlap together and are shown by the black bold lines.
The other parameters are the same as those of Fig.2. }
\end{figure}

Next we study the transport properties of the graphene/superconductor hybrid system
in the regime of high magnetic fields.
In Fig.4, we plot the Andreev reflection coefficients $T_{nmA}$ and the normal transmission coefficients
$T_{13N}$ versus the incident energy $\omega$ for different strong magnetic fields.
The $T_{11A}$, $T_{33A}$, and $T_{31A}$ are completely suppressed and are almost zero, and only $T_{13A}$ and $T_{13N}$ are remarkable.
In addition, $T_{13A}+T_{13N}=1$ when $\omega$ is within the superconductor gap $\Delta$.
These are similar to the results of $\phi =0.003$ in Fig.3.
When the bias of the terminal-1 is equal to that of the terminal-3, i.e., $V_1=V_3$,
the normal tunneling term $T_{13N}$ does not contribute to the current (see Eq.(9)).
In the case of small bias ($|eV|<\Delta$), the tunneling from the graphene to the superconductor is also
prohibited ($T_{23N}=0$) due to the existence of the superconductor gap.
Then, the current in Eq.(9) can be reduced to
\begin{equation}
I_{3}=(2e/h)\int{\rm d}\omega \{\ T_{13A}(f_{3+}-f_{1-})+T_{33A}(f_{3+}-f_{3-}) \}\ ,
\end{equation}
in which only the Andreev reflection contributes to the current.
Because $T_{11A}=T_{33A}=0$ at strong magnetic field, the differential conductance is
$G_3 = \frac{d I_3}{dV} =\frac{2e^2}{h}[ T_{13A}(eV)+T_{13A}(-eV)]$ at zero temperature.
Thus, in the regime of strong magnetic field, an exclusive CAR $T_{13A}$ is obtained, with both the LAR and the normal tunnelling being completely prohibited.
Notice that since the relation $T^{\uparrow}_{13A}= T^{\downarrow} _{13A}$ always holds, and the spin-up and spin-down electrons
from a Cooper pair transport to the terminal-1 and the terminal-3 randomly, the two spatially separated electrons
can keep their spin and momentum entangled.
By setting the bias of the superconductor slightly higher than one of the graphene terminals,
the current can be driven from the superconductor to the graphene,
and the Cooper pair can be split into two separated electrons which will flow into different leads (see Fig.1(a)).
This generates two spatially separated electron flows with entangled spin and momentum.

Here, we emphasize that the efficiency of this Cooper-pair splitter is very high,
although the value of the CAR shows oscillation behavior.
From Eq.(10), we can define splitting efficiency as:
\begin{eqnarray}
\eta =\frac{\int{\rm d}\omega \{T_{13A}(f_{3+}-f_{1-}) \}}
{\int{\rm d}\omega \{ T_{13A}(f_{3+}-f_{1-})+T_{33A}(f_{3+}-f_{3-}) \}}.
\end{eqnarray}
Here, $\eta$ describes the probability to obtain two spatially separated electrons
when a Cooper pair is split.
Because the LAR is completely suppressed with $T_{33A}=0$ in the quantum Hall regime,
the splitting efficiency $\eta$ of a Cooper pair is always $100\%$.

\begin{figure}
\includegraphics[width=8.7cm,totalheight=6.5cm, clip=]{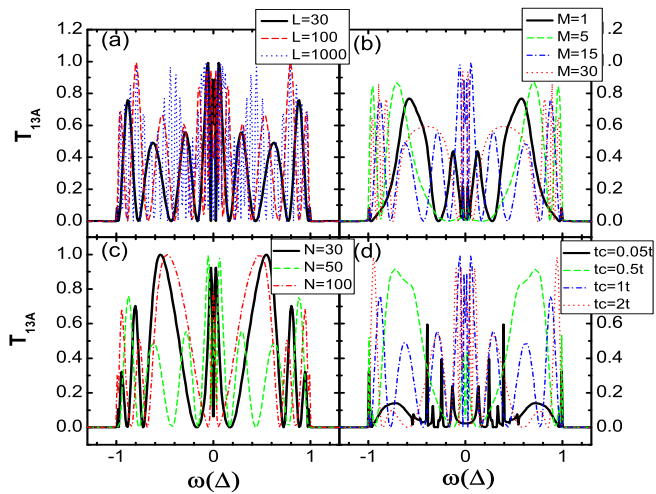}
\caption{(Color online)
Energy-dependent CAR coefficient $T_{13A}$ by considering different $L$ (a), $M$ (b), $N$ (c), and $t_c$ (d) in the graphene/superconductor hybrid system.
If the parameters are not shown in the legend, they are as follows:
$L=30$, $N=50$, $M=15$, $t_c=t$, $E_0=-5\Delta$, $\phi=0.01$, and $W=0$. Here, $T_{11A}$, $T_{31A}$, and $T_{33A}$ are not shown in the figure because they are all zero. }
\end{figure}

Now let us study how the exclusive CAR coefficient $T_{13A}$ in the graphene/superconductor system
is affected by the system parameters. Fig.5(a) shows $T_{13A}$ versus $\omega$ for different length
$L$ of the central region. We find that $T_{13A}$ can always reach large value by varying $L$ from 30 to 1000.
In addition, $T_{13A}$ oscillates with the energy $\omega$, and the oscillation frequency
is increased by increasing the length $L$,\cite{resonance,resonance2}
because the Fabry-P\'{e}rot-like interference occurs in this three-terminal system.
In particular, for $L=1000$, the length of the central region is $\sqrt{3}aL\approx246$nm which is much greater than the superconducting coherent length.
However, the CAR coefficient $T_{13A}$ is still very large in this case.
As compared with previously proposed Cooper-pair splitters that the length of the central region should be less than the superconductor coherent length $\xi$, we show in the present study that the CAR process is not confined by
the length $L$, because of the unidirectional chiral edge states.
The exclusive CAR coefficient $T_{13A}$ can also be quite large by changing the width $N$ of the graphene nanoribbon (as shown in Fig.5(c)).
In fact, as long as the graphene nanoribbon is wide enough so that the top and bottom chiral edge states don't mix together, an exclusive CAR process can always be obtained.
Figs.5(b) and 5(d) show how $T_{13A}$ be affected by the covered area and the coupling strength
between the superconductor and the graphene. By varying the width $M$ of the covered area of the superconductor on the graphene
and the coupling strength $t_c$, $T_{13A}$ can still have large value.
These results show that the exclusive CAR process in our system can hold very well in a wide range of
the system parameters, which is helpful for experimental implementation of the Cooper-pair splitter.

\begin{figure}
\includegraphics[width=8.7cm,totalheight=6cm, clip=]{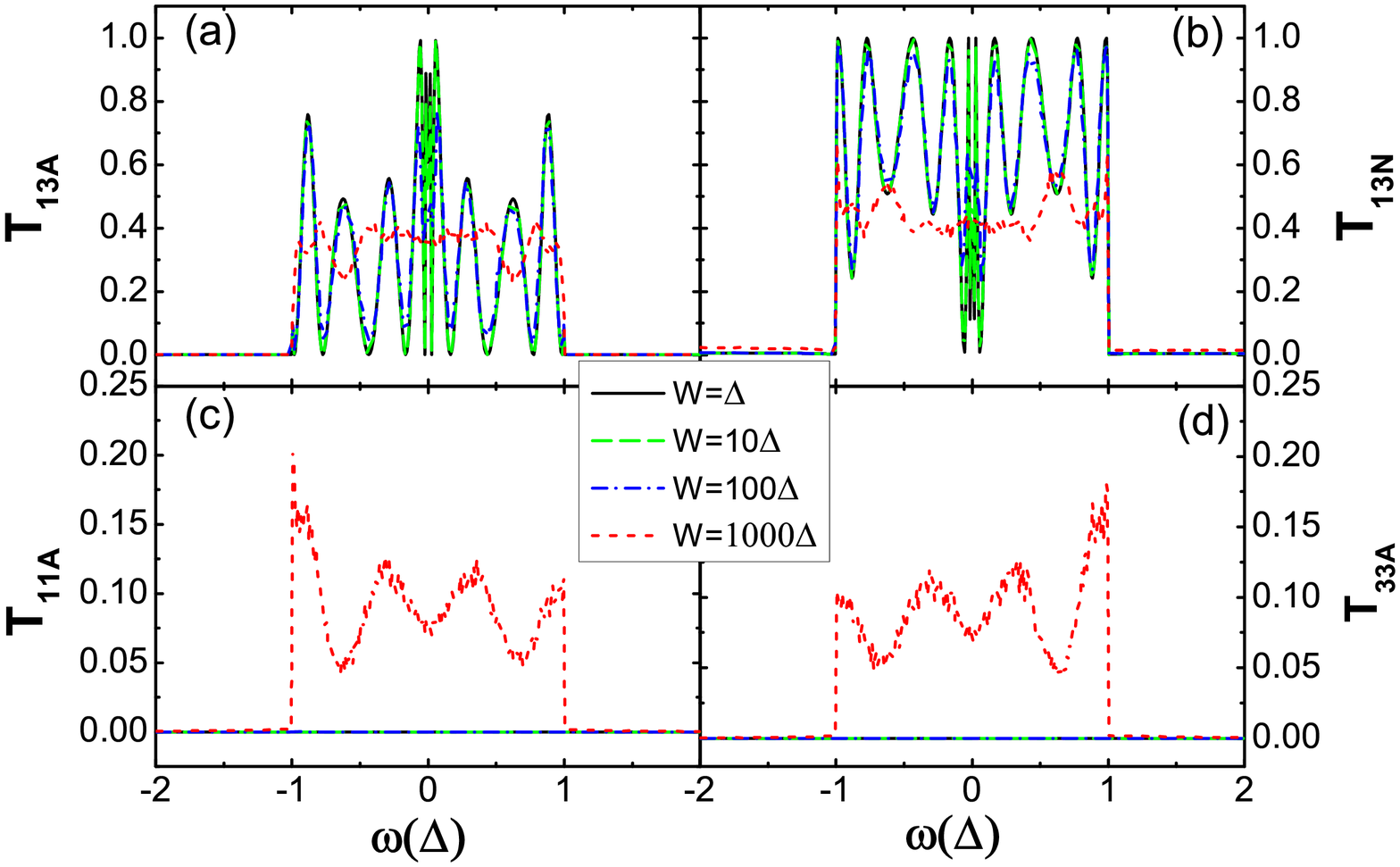}
\caption{ (Color online)
$T_{A}$ and $T_{N}$ versus $\omega$ for different disorder strengths $W$. The parameters are $E_0=-5\Delta, N=50, L=30, M=15, t_c=t$, and $\phi=0.01$. }
\end{figure}

Let us study the effect of the disorder on the exclusive CAR. Here, we consider the on-site Anderson
disorder which only exists in the central region.
Fig.6 shows the Andreev reflection coefficients $T_A$ and the normal transmission coefficient $T_{N}$
versus the incident energy $\omega$ for different disorder strengths $W$. By increasing the disorder strength $W$ from 1meV to 100meV, the above coefficients are almost unaffected,
indicating that our Cooper-pair splitter is very robust. By further increasing $W$ to 1000meV, the LAR coefficients
$T_{11A}$ and $T_{33A}$ become nonzero, and the oscillation behavior of $T_{13A}$ and $T_{13N}$ disappears.
This is due to the fact that the system goes into the diffusive regime at large disorder and the edge states are destroyed by the disorder. Therefore, as long as the edge states survive, the exclusive CAR $T_{13A}$ can persist. This means that
the exclusive CAR, i.e., the Cooper-pair splitter, is robust owing to the quantum Hall effect.

In Appendix, we show the electron transport properties of a three-terminal 2D electron gas/superconductor hybrid
system and obtain similar results as the graphene/superconductor hybrid system. Although the two systems have
different band structures and electronic behaviors, both of them can work as a perfect Cooper-pair splitter, owing to the chiral edge states which are induced by strong magnetic field.
However, in general, the graphene/superconductor
system has more advantages as compared with the 2D electron gas/superconductor one.
This is attributed to the fact that the graphene has a unique band structure with a linear dispersion
relation near the Dirac point, and the zeroth Landau level at the Dirac point is well separated
from the first Landau level.

\begin{figure}
\includegraphics[width=8.7cm,totalheight=7.5cm, clip=]{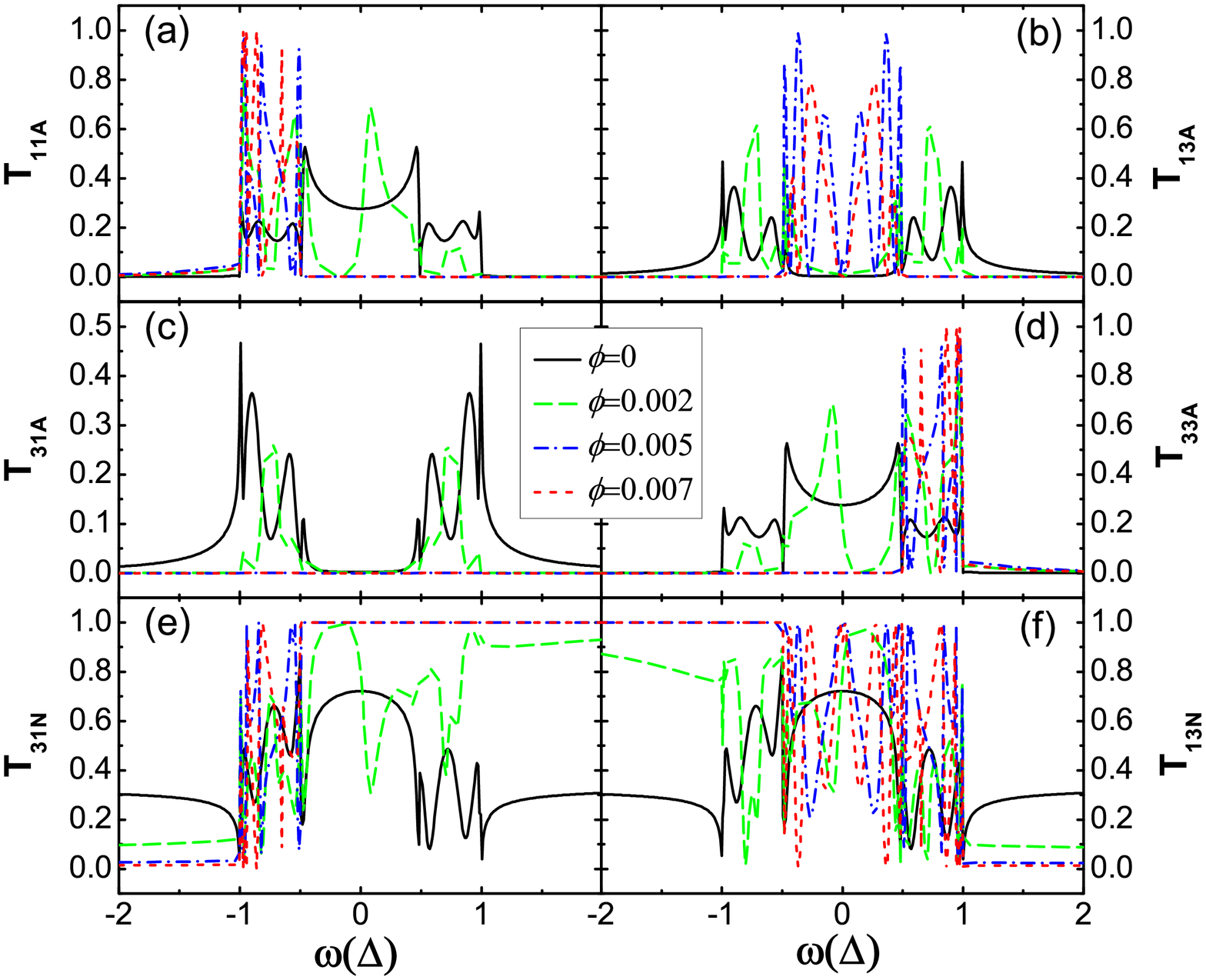}
\caption{ (Color online)
Andreev reflection coefficients $T_A$ and normal transmission coefficients $T_{N}$
versus energy $\omega$ for the graphene/superconductor hybrid system under different magnetic fields, with the Dirac point energy $E_0=-0.5\Delta$. The other parameters are $N=50$, $L=30$, $M=15$, $t_c=t$, and $W=0$.}
\end{figure}

\section{GRAPHENE/SUPERCONDUCTOR SYSTEM with Dirac point near Fermi energy}

Let us change the Dirac point energy $E_0$ from $-5\Delta$ to $-0.5\Delta$ to see the interesting behavior of
the LAR and the CAR in the graphene/superconductor system. In the experiment, the Dirac point energy can be easily tuned by the gate voltage.
At zero magnetic field, the Andreev reflection in the graphene/superconductor interface
can be divided into the retro-reflection and the specular reflection, according to the situation that the incoming electron and the reflected
hole locate in the same bands or different ones (see Figs.8(b) and 8(c)).\cite{CAR4,Chengshuguang,XYXnew}
When the Dirac point is near the Fermi energy, the specular reflection is dominant.
While in the presence of the magnetic field, the movement of the reflected hole will be changed and the results are totally different.
Fig.7 shows the Andreev reflection coefficients $T_A$ and the normal transmission coefficients $T_{N}$ versus
the incident electron energy $\omega$ under different magnetic fields. Similar to the case of $E_0=-5\Delta$, the edge states gradually form in the three-terminal system by increasing the magnetic field and some unique properties appear. Here, we emphasize the following two facts:
(i) When $|\omega|>|E_0|=0.5\Delta$, the incident electron and the reflected hole locate in different bands, i.e., the valence band ($E<E_0$) with negative chirality and the conduction band ($E>E_0$) with positive chirality (see Fig.8(b)). Thus, under strong magnetic field, the direction of the unidirectional chiral edge states is opposite for the incident electron and the reflected hole.
(ii) When $|\omega|<|E_0|$, the incident electron and the reflected hole locate in the same band (see Fig.8(c)), and the direction of the chiral edge states is the same.

Now we focus on the case of high magnetic field (e.g., $\phi=0.007$) in Fig.7.
For the incident electron from the terminal-1, the electron transports along the top edge state from the left to the right when $\omega<E_0=-0.5\Delta$ (see Fig.8(a)), and then it meets the superconductor and will be Andreev reflected. The reflected hole transports along the opposite direction as compared with the incident electron.
As a result, $T_{11A}$ is considerably large and $T_{31A}=0$ (see Figs.7(a) and 7(c)). In this case, the two electrons in the Cooper pair get into the same terminal (see Fig.8(a)).
When the incident energy $\omega$ is above $E_0$, the electron goes along the bottom edge state from the left to the right
and does not encounter the superconductor. Then, $T_{11A}=T_{31A}=0$ and $T_{31N}=1$, as shown in Figs.7(a), 7(c), and 7(e).
Similar results can be obtained for the incident electron from the terminal-3. When $\omega<E_0$, the electron goes along the bottom edge state from the right to the left without meeting the superconductor, and $T_{13A}=T_{33A}=0$ and $T_{13N}=1$ (see Figs.7(b), 7(d), and 7(f)); When $E_0<\omega<-E_0$ (see Fig.8(c)), both the incident electron and the reflected hole are in the conduction band, and go along the top edge states from the right to the left. This corresponds to the exclusive CAR process and $T_{13A}$ has large value with $T_{33A}=0$ (see Figs.7(b) and 7(d)). When $-E_0<\omega$, the incident electron and the reflected hole are in different bands, and only the LAR happens in the graphene/superconductor interface. Note that in Figs.7(a)-7(d), the LAR and the CAR can be separated completely by tuning the bias. Therefore, this provides us a good way to control and investigate these processes.

\begin{figure}
\includegraphics[width=4.3cm,totalheight=2.3cm, clip=]{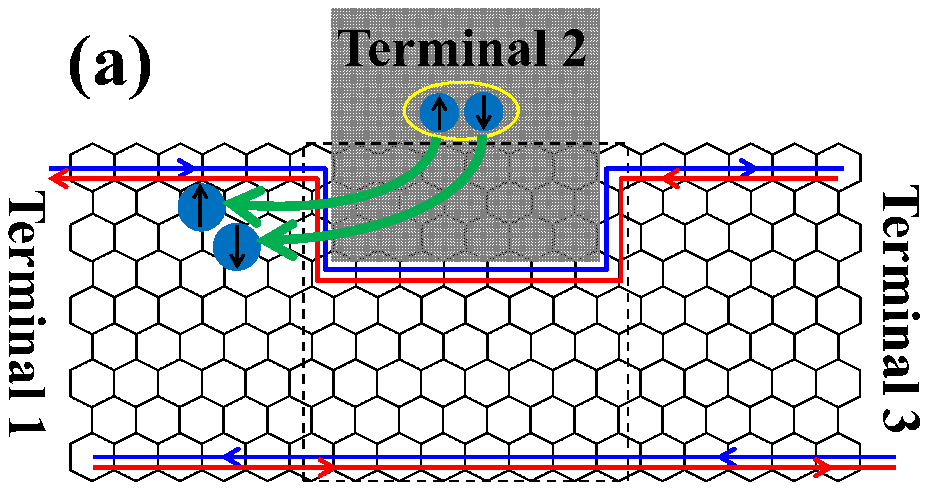}
\includegraphics[width=4.1cm,totalheight=2.3cm, clip=]{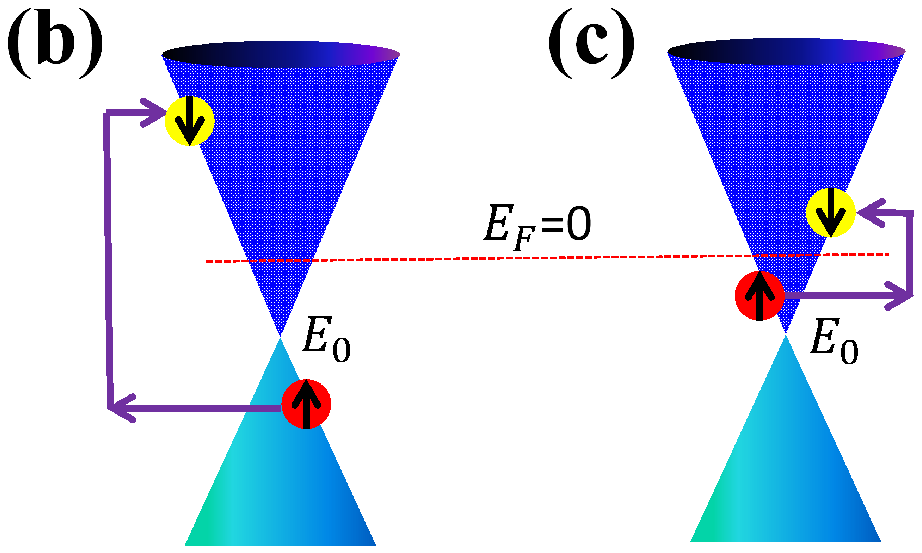}
\caption{ (Color online) (a) Schematic diagram of positive chiral edge states (red lines) for the conduction band and
negative chiral edge states (blue lines) for the valence band in the three-terminal graphene/superconductor system.
(b) and (c) Schematic view of the specular reflection and the retro-reflection in the energy space.
While in the presence of the magnetic field and the chiral edge states,
the specular reflection and the retro-reflection are changed into the LAR and the CAR, respectively.
}
\end{figure}

\section{conclusion}

In summary, we investigate the electron transport in a three-terminal graphene/superconductor hybrid system.
In high magnetic field, an exclusive crossed Andreev reflection is obtained with the aid of the edge states in the quantum Hall regime, with the other Andreev reflections being
prohibited. This exclusive crossed Andreev reflection can hold by varying the size of the system and the coupling strength
between the graphene and the superconductor.
In particular, it can also work well for large width of the central superconductor and is very robust against the disorder.
As a result, a perfect Cooper-pair splitter with high efficiency is proposed in this study. Finally, a two-dimensional electron gas/superconductor quantum Hall system is also considered,
where similar results are obtained due to the chiral edge states in the system.

\section*{\bf ACKNOWLEDGMENTS}

We gratefully acknowledge the financial support from NBRP of China (2012CB921303 and 2015CB921102),
NSF-China under Grants No. 11274364, 11504066, and 11574007, and FRFCU under Grant No. AUGA5710013615.
Y. Xing is supported by the grant of Trans-Century Training Programme Foundation for the Talents by
the State Education Commission (No. NCET-13-0048).

\section*{APPENDIX}

In this appendix, we provide the electron transport properties of a three-terminal system, which is composed of a 2D electron gas nanoribbon and a superconductor lead (see Fig.1(b)). In the tight-binding representation, the Hamiltonian of the 2D electron gas nanoribbon is:
\begin{eqnarray}
H_{\rm EG}=\sum_{i,\sigma}E_ia_{i\sigma}^{\dagger}a_{i\sigma}-
\sum_{\langle ij\rangle,\sigma}t{\rm e}^{{\rm i}\phi_{ij}}a_{i\sigma}^{\dagger}a_{j\sigma}.
\end{eqnarray}
Here, $t=\frac{\hbar^2}{2ma^2}$ is the kinetic energy, $E_i=E_b+4t+w_i$ with $E_b$ the bottom of the conduction band, and the magnetic field is described by the magnetic flux $\phi$ in a square lattice. The Hamiltonians of the superconductor and the coupling between the superconductor and the electron gas are the same as Eqs.(3) and (4), respectively. Then, the Andreev reflection coefficients $T_A$ and the normal transmission coefficients $T_N$ can be calculated from Eqs.(5) and (6). In the numerical calculation, we set the superconductor gap $\Delta=t/200$, the conduction band bottom $E_b=-0.02t$, and the Fermi energy of the superconductor $E_F=0$.

\begin{figure}
\includegraphics[width=8.7cm,totalheight=7.5cm, clip=]{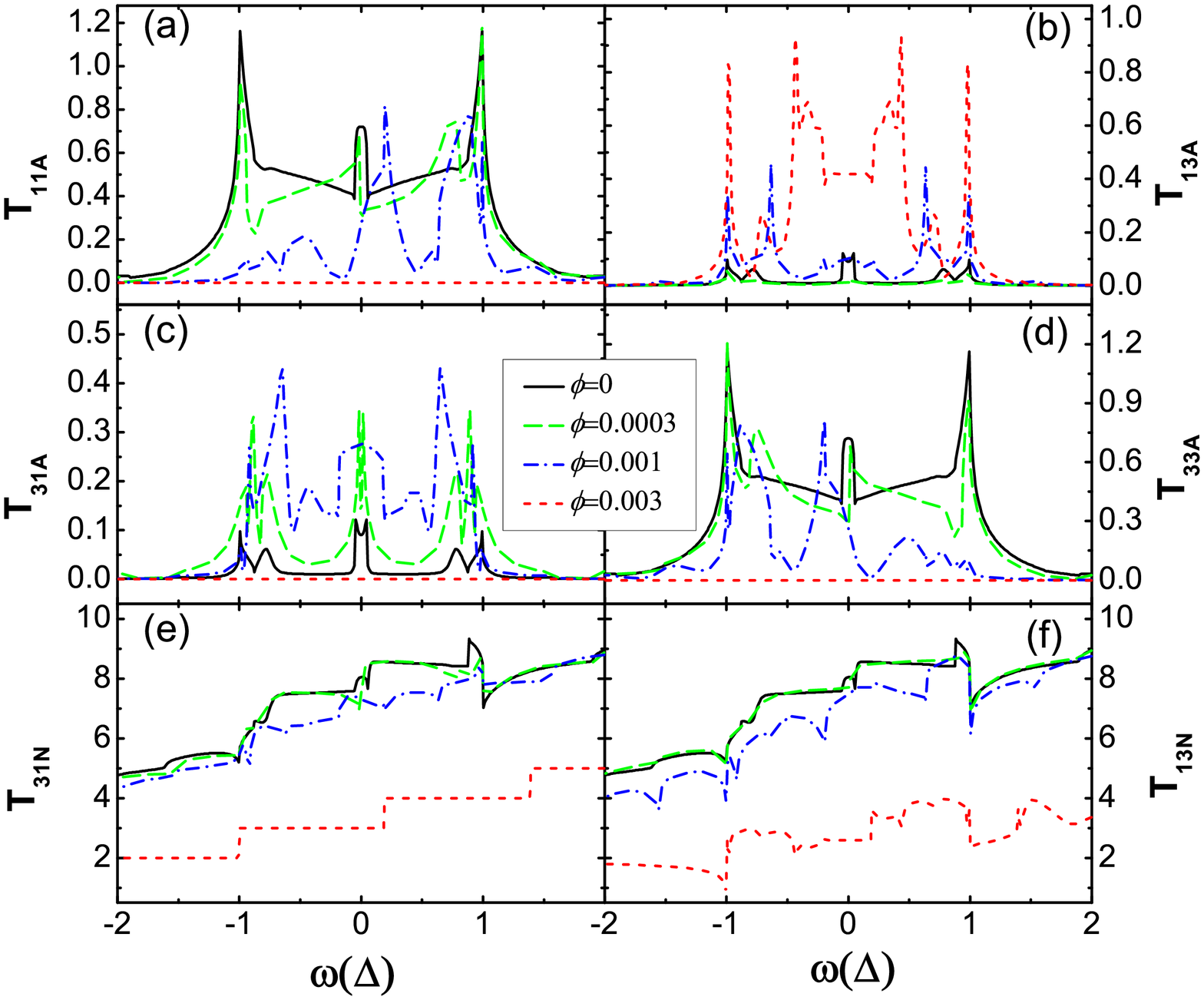}
\caption{ (Color online) Andreev reflection coefficients $T_A$ and normal transmission coefficients $T_{N}$ versus incident energy $\omega$ for a 2D electron gas/superconductor system under different magnetic fields. The parameters are $E_b=-0.02t$, $N=100$, $L=50$, $M=20$, $t_c=0.03t$, and $W=0$.}
\end{figure}

\begin{figure}
\includegraphics[width=8.7cm,totalheight=6cm, clip=]{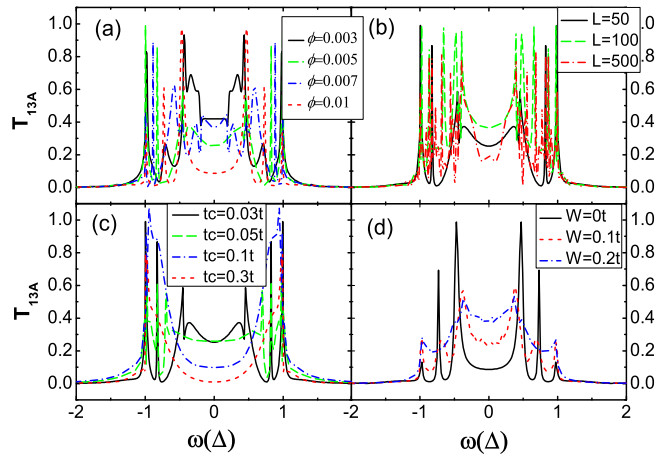}
\caption{ (Color online)
$T_{13A}$ versus $\omega$ for different $\phi$ (a), $L$ (b), $t_c$ (c), and $W$ (d)
in the electron gas/superconductor hybrid system with $E_b=-0.02t$, $N=100$, and $M=20$.
The other parameters are: (a) $L=50$, $t_c=0.03t$, and $W=0$;
(b) $t_c=0.03t$, $\phi=0.005$, and $W=0$; (c) $L=50$, $\phi=0.005$, and $W=0$; and
(d) $L=50$, $t_c=0.03t$, and $\phi=0.01$.
$T_{11A}$, $T_{31A}$, and $T_{33A}$ are almost zero so they are not shown in the figure.}
\end{figure}

Fig.9 shows the transport properties under different magnetic fields. In zero or weak magnetic field, both the LAR process and the CAR one occur (see Figs.9(a)-9(d)). By increasing the magnetic field to $\phi=0.003$, the edge states gradually form just as the graphene based hybrid system.
Since $\phi =a^2 B/\phi_{0}$, the corresponding magnetic field $B$ is about 0.078 Tesla for $\phi=0.003$ and $a=5$nm.
Due to the edge states, an exclusive CAR $T_{13A}$ can be obtained, with the other Andreev reflection coefficients $T_{11A}$, $T_{31A}$, and $T_{33A}$ being zero [see Figs.9(a)-9(d)]. In addition, some Hall plateaus emerge in the curve of $T_{31N}$ [see Fig.9(e)], where the plateau values are determined by the filling factor $\nu$ of the Landau levels. This indicates that the incident electron from the terminal-1
tunnels directly into the terminal-3 without the interface scattering.
With the aid of the edge states, these results can also be well understood.
Now both the incident electron and the reflected hole are in the conduction band, and move anticlockwise along the edge of the electron gas under high magnetic field, as shown in Fig.1(b). For the incident electron from the terminal-3, both the CAR $T_{13A}$ and the direct tunneling $T_{13N}$ occur. For the incident electron from the terminal-1, only the direct tunneling $T_{31N}$ occurs.
The LAR is completely prohibited, regardless of the terminal where the electron is injected. Therefore, the Cooper-pair splitter is also very efficient in the 2D electron gas/superconductor system. In fact, as long as the edge states form, one can always demonstrate a perfect Cooper-pair splitter in such quantum Hall systems.

Finally, we study the exclusive CAR coefficient $T_{13A}$ by considering the influence of the magnetic field $\phi$, the length $L$ of the central region, and the coupling strength $t_c$ between the superconductor and the electron gas. Figs.10(a)-10(c) show $T_{13A}$ under large magnetic field, where the Hall edge states emerge in the system. It can be seen that $T_{13A}$ is quite large for a wide range of the system parameters, and the other Andreev reflection coefficients $T_{11A}$, $T_{31A}$, and $T_{33A}$ are almost zero. Thus, the proposed Cooper-pair splitter based on the quantum Hall chiral edge states can work well for the large width
of the central superconductor and for a wide range of the system parameters. We also consider the situation when the on-site energy disorder is introduced in the central region, as illustrated in Fig.10(d). It is evident that $T_{13A}$ is very robust against the disorder. These results are similar to the case of the graphene/superconductor system.

\end{document}